\pdfoutput=1
\documentclass[11pt]{article}
\usepackage{pdfpages}
\usepackage{graphicx}
\usepackage{amssymb}
\usepackage{amsmath}
\usepackage{amsfonts}
\usepackage{enumerate}
\usepackage{amsthm}
\usepackage{moreverb}
\usepackage[colorlinks,bookmarksopen,bookmarksnumbered,citecolor=red,urlcolor=red]{hyperref}
\usepackage{mathtools}
\usepackage{stackengine}
\usepackage{scalerel}
\usepackage{multirow}
\usepackage{natbib}

\begin{document}

\title{{\Large Interval estimators for inequality measures using grouped data}
\author{Dilanka S. Dedduwakumara, Luke A. Prendergast \\
Department of Mathematics and Statistics\\ 
 La Trobe University
 }
 } 
\date{} 
\maketitle

\begin{center}
\textbf{Abstract}
\end{center}

Income inequality measures are often used as an indication of economic health. How to obtain reliable confidence intervals for these measures based on sampled data has been studied extensively in recent years. To preserve confidentiality, income data is often made available in summary form only (i.e. histograms, frequencies between quintiles, etc.). In this paper, we show that good coverage can be achieved for bootstrap and Wald-type intervals for quantile-based measures when only grouped (binned) data are available.  These coverages are typically superior to those that we have been able to achieve for intervals for popular measures such as the Gini index in this grouped data setting.  To facilitate the bootstrapping, we use the Generalized Lambda Distribution and also a linear interpolation approximation method to approximate the underlying density.  The latter is possible when groups means are available.  We also apply our methods to real data sets.

\vspace*{.3in}

\newpage

%
%
%
\section{Introduction} \label{Introduction}
 
Income data are generally made available in binned formats by governing bodies to preserve the confidentiality of the individual participants. Obtaining inferences from such summary information has been recently discussed by \cite{doi:10.1080/03610918.2018.1499935}, in the context of obtaining confidence intervals for quantiles using estimates of the underlying distribution using grouped data. As we will show in what follows, we can obtain reliable confidence intervals for some inequality measures using bootstrap and Wald-type approaches. 

Motivated by these findings, we compare the interval estimators for inequality measures when the data are available in grouped form only. For comparison, we use the well-known Gini, Theil and Atkinson indices and the newly proposed quantile ratio index \citep{prendergast2016simple}.  We begin by introducing these measures before discussing some distribution estimation strategies in Section 3.  In Section 4, we report findings of simulations for interval estimators of the inequality measures.  Two real data examples are presented in Section 5, followed by a brief discussion in Section 6.

\section{Some inequality measures} \label{Inequality Measures}

Let $f$, $F$ and $Q$ denote the density, distribution and quantile functions respectively for the population of interest.  For $p \in [0,1]$, let $x_p=Q(p)=F^{-1}(p)$ denote the $p$-th quantile.  We find it convenient to consider continuous probability distributions to model incomes while acknowledging that, in practice, a population of incomes has a finite number, $N$, of individuals.   Let $x_1,\ldots,x_n$ denote a simple random sample of incomes from the population and let $\widehat{x}_p$ be the estimated $p$-th quantile.  

\subsection{Gini Index} \label{Gini Index}

Suppose $X\sim F$ where $X$ represents a randomly chosen income from the population and let $\mu=E(X)$ denote mean income.  Easily the most commonly used inequality measure is the Gini index \citep{gini1914sulla}, which measures the deviation of the income distribution from perfect equality. It can be defined as, 
\begin{equation*}
G =1-\frac{1}{\mu}\int\limits_0^\infty [1-F(x)]^2 \ dx 
\label{eqn:eq2}
\end{equation*}
with $G \in [0,1]$.  Here, $G=1$ indicates that one individual holds all wealth (e.g. one individual with income greater than zero) and $G=0$ represents the equality of incomes for all.
The Gini index can be estimated for a simple random sample of size $n$, with the ordered values of $x_1,\ldots,x_n$ by, 
\begin{equation*}
\hat{G}=\frac{2\sum_i i x_i}{n\sum_i x_i}-\frac{n+1}{n}.
\label{eqn:eq2}
\end{equation*}
For more details on the Gini index and estimation see, for example, \cite{dixon1988erratum} and \cite{damgaard2000describing}.

\subsection{Theil index} \label{THeil Index}
Based on information theory, \cite{theil1967economics} proposed an entropy-based measure which is defined to be  
\begin{equation*}
T =\int\limits_0^\infty \Big(\frac{x}{\mu} \Big)\log\Big(\frac{x}{\mu}\Big) f(x)\ dx 
\label{eqn:eq2}
\end{equation*}
where $T \in [0,\infty)$.  In practice where a population consists of finite number of $N$ incomes, the upper bound is $\ln(N)$. The Theil index can be estimated by
\begin{equation*}
\widehat{T} = \frac{1}{n} \sum_i\frac{x_i}{\bar{x}}\ln\bigg(\frac{x_i}{\bar{x}}\bigg) 
\label{eqn:eq2}
\end{equation*}
where $\bar{x}$ is the sample mean and where $\widehat{T}\in [0, \ln(n)]$.  Further properties of the Theil index can be found in \cite{theil1967economics}, \cite{allison1978measures} and \cite{shorrocks1980class}.

\subsection{Atkinson index} \label{Atkinson Index}

The Atkinson index was initially introduced by \cite{atkinson1970measurement}. This measure depends on the sensitivity parameter, $\epsilon$ $(0<\epsilon < \infty)$, which represents the level of inequality aversion. As this parameter increases, more weight is shifted to the distribution at the lower end and vice versa. It is defined as
\begin{equation*}
A =1-\left[ \int\limits_0^\infty \Big(\frac{x}{\mu} \Big)^{1-\epsilon}f(x) \ dx \right]^{\frac{1}{1-\epsilon}}
\label{eqn:eq2}
\end{equation*}
where $ A \in [0,1]$.  

Atkinson values represent the proportion of total income that would be needed to achieve an equal level of social welfare if incomes were perfectly distributed. Depending on the value of $\epsilon$, the sample estimate is
\begin{equation*}
\hat{A} =
\begin{cases}
\displaystyle 1-\frac{1}{\bar{x}}\bigg(\frac{1}{n}\sum_i x_i^{1-\epsilon} \bigg)^{\frac{1}{1-\epsilon}}, & \text{for} \quad 0 \leq \epsilon < 1  \\
\displaystyle 1-\frac{1}{\bar{x}}\bigg(\prod_i x_i\bigg)^{\frac{1}{n}}, & \text{for} \quad \epsilon = 1  
\end{cases}
\label{eqn:eq2}
\end{equation*}

We use the value of $\epsilon=0.5$ for our analysis which is the default value used in the package \texttt{ineq} \citep{ineq} in \texttt{R} software \citep{R}. More details for the Atkinson index can be found in \cite{atkinson1970measurement}, \cite{biewen2006variance} and \cite{shorrocks1980class}.

\subsection{Quantile ratio index} \label{Quantile ratio index}
\cite{prendergast2016simple, pr&st2019} introduced the quantile ratio index (QRI) which uses the ratio of symmetric quantiles and which is simpler than similarly defined inequality measures given by \cite{prendergast2016quantile}. The QRI is denoted as 
\begin{equation*}
I = 1-\int\limits_0^1 \frac{x_{p/2}}{x_{1-p/2}} \ dp =1-\int\limits_0^1 R(p) \ dp
\label{eqn:eq1}
\end{equation*}
where $I \in [0,1]$.  Note that $R(p)$ is the ratio of symmetric quantiles so that $I$ can be seen to be based on the average ratio of incomes chosen symmetrically from the poorer and richer halves of the incomes respectively.  For a suitably large $J$, $I$ is estimated as $J^{-1}\sum_j\left[1-\widehat{R}(p_j)\right]$ where $p_j=(j-1/2)/J$ and $\widehat{R}(p_j)$ is the ratio of the estimated $(p_j/2)$-th and $(1-p_j/2)$-th quantiles.  \cite{prendergast2016simple} show that $J=100$ is large enough to obtain good estimates of $I$ and so this will be our choice in what follows.

\section{Density estimation methods} \label{Methods}

We now consider two methods for estimating the density from grouped data.  The first requires bins and frequencies, and the second also requires the bin means.  The methods were used by \cite{doi:10.1080/03610918.2018.1499935} to obtain intervals for quantiles from histograms. 

\subsection{GLD Estimation method} \label{GLD Estimation method}

Due to flexibility in approximating a wide range of distributions, the Generalized Lambda Distribution (GLD) is commonly used and particularly favoured in fields such as economics and finance. Defined in terms of its quantile function, several parameterizations for the GLD exist. Following is the FKML parameterization for the GLD given by \cite{freimer1988study} which is often favoured since it is defined for all parameter choices, with the only restriction being that the scale parameter must be greater than zero.  The GLD quantile function is 
\begin{equation}
Q(p)=\lambda + \frac{1}{\eta} \bigg[\frac{(p^\alpha-1)}{\alpha}-\frac{(1-p)^\beta-1}{\beta}\bigg].
\label{eq1}
\end{equation}

The GLD has been used in different contexts to obtain various interval estimators \citep[e.g.][]{su2009confidence,prendergast2016exploiting} when the full data set is available. However, using the percentile matching methods presented by \cite{karian1999fitting} and \cite{tarsitano2005estimation}, the GLD parameters can still be estimated when data is in grouped format with frequencies and bins. This method is available in the \texttt{bda} package \citep{wang2015bda}.

\subsection{Linear interpolation method } \label{linear interpolation method}

The linear interpolation method was proposed by \cite{lyon2016advantages} as a method of estimating the underlying distribution of binned data when the group (bin) means are also available. Within each bin, a linear density is estimated using the lower and upper bounds of the bin and the associated mean, and the final bin is fitted with an unbounded exponential tail. The slope of the linear density is determined by the mean in relation to the bin midpoint. Closed form solutions for the density and the quantile functions are extensively provided by \cite{lyon2016advantages} and following is a summary of the density results. 

Assume there are $J$ intervals in the grouped data bounded by $[a_{j-1},a_j), j=1,\ldots,J$ where $a_0>-\infty$ and $a_J=\infty$. Let the midpoint, mean and relative frequency of the $j$th bin be denoted by $x^c_j$, $\bar{x}_j$  and $\widehat{f}_j$.  The linear density for the $j$th bin is 
\begin{equation}
h_j(x)=\alpha_j+\beta_j x, \qquad x\in [a_{j-1},a_j)
\label{eq2}
\end{equation}
where the estimates of $\alpha_j$, $\beta_j$ are given by,
\begin{equation}
\widehat{\beta_j}=\widehat{f_j}\frac{12(\bar{x}_j-x^c_j)}{(a_j-a_{j-1})^3}, \quad 
\widehat{\alpha}_j=\frac{\widehat{f_j}}{a_j-a_{j-1}}-\widehat{\beta_j}x^c_j.
\label{eq3}
\end{equation}

The density estimate for the final unbounded interval using an exponential tail is provided by,
\begin{equation}
h_J(x)=\frac{\eta}{\lambda}\exp\left\{-\frac{(x-a_{J-1})}{\lambda}\right\}
\label{eq4}
\end{equation}
where $\widehat{\eta}=\widehat{f_J}$ and $\widehat{\lambda}=\bar{x}_J-a_{J-1}$.

\section{Interval estimators using grouped data}\label{CI estimation}

In this section, we propose and describe our bootstrap and Wald-type methods to produce intervals for inequality measures using grouped information. The variance of the QRI estimator depends on the underlying income distribution density function applied to income quantiles \citep{prendergast2016simple}.  Therefore, provided we can obtain good estimates of the density from grouped data, then the QRI is well-suited to obtaining Wald-type intervals in this setting.  Aside from bootstrapping, to obtain the variance of, for example, the Gini index, it is common to use the jackknife approach or other methods that require the full data set.  Consequently, obtaining an approximation to the variances for the Gini, Thiel and Atkinson measure estimators from grouped data is not straightforward and therefore an area for further research.

For the bootstrapping procedure, we obtain the bootstrap samples from the estimated quantile function arising from the estimated GLD or linear interpolation densities.  We then use the percentile bootstrap interval described below.  While there are other bootstrap methods available that often have improved performance over the percentile method, they require the full data set and it is not immediately clear on how to use them when data is only available in grouped format; e.g. the bootstrap $t$ interval requires the variance of the estimator, the BCa method \citep{efron1987better} and Efron's ABC method \citep{diciccio1992more} requires the full sample data to calculate the acceleration parameter. However, we did try a variation of the bootstrap $t$ interval whereby the $\alpha$ parameter was estimated as usual, but where the estimate and its standard error were also approximated from the bootstrap samples given the lack of the full data set.  Coverages were usually no better, and often worse than those for the percentile approach so we do not present them in what follows for brevity.  Further variations of bootstrap methods to accommodate the lack of the full data set may result in improved results and this is an area for future research.

\subsubsection{Bootstrap confidence intervals.} \label{Simulations} 
In the following algorithm, we describe the estimation of percentile bootstrap confidence intervals in detail.

\begin{enumerate}[Step 1:]
            \item Estimate the GLD and linear interpolation densities using available summary information of bin points and frequencies (and bin means for the linear interpolation approach).
            \item Take 500 bootstrap samples of size \textit{n} using the estimated quantile functions from the two estimation methods using the inverse transform sampling method. That is, randomly generate $n$ numbers, $y_1,\ldots, y_n$ in $[0,1]$ from the uniform distribution and then the $i$th observation for the $j$th bootstrap is $y_{ji}=\widehat{Q}(y_i)$ where $\widehat{Q}$ is the estimated quantile function.
            \item Construct the percentile bootstrap 95\% confidence intervals by taking the 2.5\% and 97.5\% quantiles of the 500 bootstrapped estimates of the inequality measures.
         \end{enumerate}

For the GLD method, we consider the available bin points as the empirical percentiles in the percentile matching method, providing the estimated parameters for the GLD. By using the GLD quantile function (Section \ref{GLD Estimation method}) and the estimated parameters, we can easily take the bootstrap samples using the inverse transform sampling method as in Step 2. For the linear interpolation approach, we use the following two quantile functions to generate data depending on the value of \textit{p} \citep{lyon2016advantages}.   For the bounded interval of $[a_{j-1},a_j)$, the following quantile function is used for $p\in [0,1)$ is,
\begin{equation}
\widehat{x}_p=\frac{-\widehat{\alpha}_j + \sqrt[]{2\widehat{\beta}_jp+\widehat{C}_j}}{\widehat{\beta}_j}
\label{eq5}
\end{equation}
where, $\widehat{C_j}=[\widehat{\alpha}_j^2-2\widehat{\beta}_j\widehat{F}_{j-1}+2\widehat{\beta}_j\widehat{\alpha}_j a_{j-1}+\widehat{\beta}_j^2(a_{j-1})^2]$, $\widehat{\beta_j}$ and $\widehat{\alpha_j}$ as in \eqref{eq3}.

Further the fitted exponential tail yields the following quantile function when the cumulative relative frequency up to final (\textit{J}th) interval is denoted  by $\widehat{F}_J$,
\begin{equation}
\widehat{x}_p=a_{J-1}-\widehat{\lambda}\ln\Bigg(1-\frac{p-\widehat{F}_{J-1}}{\widehat{\eta}}\Bigg).
\label{eq6}
\end{equation}

\subsubsection{Wald-type  confidence intervals for the QRI.} \label{Simulations} 

Obtaining confidence intervals for the QRI from full data sets is studied by \cite{prendergast2016simple}.   The variance of the estimator depends on the density function and quantiles.  Therefore, given a good estimation of the density which in turn would be expected to give good estimates to quantiles, QRI intervals from grouped data are possible.

The $(1-\alpha)\times 100$ confidence interval for \textit{I} is given by $\hat{I} \pm z_{1-\alpha/2}\sqrt[]{\text{Var}(\hat{I})}$, where $\text{Var}(\hat{I})$ is adopted from \cite{prendergast2016simple} where we use $J=100$. Here, $z_{1-\alpha/2}$ is the $1-\alpha/2$ percentile from the standard normal distribution. $\text{Var}(\hat{I})$ consists of the variances and co-variances terms of ratios of symmetrically chosen quantiles \citep[see][]{prendergast2016simple}. We then require estimates for population quantiles and density function. As described earlier, first we estimate the underlying density and quantile functions using the GLD and linear interpolation methods. Then those estimated quantile functions can be used to estimate the symmetrically chosen quantiles.

\section{Simulations and Examples} \label{Simulations and Examples} 

We begin by reporting our findings for simulation studies conducted with a variety of distributions before considering real data examples.

\subsection{Simulations} \label{Simulations} 
  
To assess coverage, we consider the lognormal distribution with $\mu=0$ and $\sigma=1$ and the Singh-Maddala distribution with parameter values $a=1.6971$, $b=87.6981$ and $q=8.3679$ where these parameters were from fitted US family incomes reported by \cite{mcdonald1984some}. We also consider the Dagum distribution with the parameter choices of $a=4.273$ $b=14.28$ and $p=0.36$ which were used in \cite{kleiber2008guide} and were estimated from fitted US family incomes in 1969.  The $\chi^2_2$, Pareto type II distribution with scale one and shape equal to two and the exponential distribution with rate one were also considered. Table \ref{tab1} provides the population inequality values of each measure.

\begin{table}
\small
  \centering
  
    \caption{True values of inequality measures for each distribution.\strut}
\label{tab1}
\hspace*{-1cm}
\begin{tabular}{|c|c|c|c|c|}
\hline
\textit{F }& \textit{Gini} & \textit{Theil} & \textit{Atkinson} & \textit{I} \\
\hline
Lognormal & 0.520 & 0.500 & 0.221 & 0.664\\
Singh-Maddala & 0.355 & 0.206 & 0.106 & 0.579 \\
Dagum & 0.335 & 0.191 & 0.097 & 0.548 \\
$\chi^2_2$ & 0.500 & 0.423 & 0.215 & 0.702 \\
Pareto(2) & 0.667 & 1.000 & 0.383 & 0.740 \\
Exponential(1)  & 0.500 & 0.423 & 0.215 & 0.702 \\
Weibull(10) & 0.067 & 0.007 & 0.004 & 0.167 \\
\hline
\end{tabular}
 \end{table}

\begin{table}
\small
\renewcommand{\tabcolsep}{1.5mm}
  \centering
  
    \caption{Empirical coverage probabilities and average widths (in brackets) of Boot-strapped interval estimates of inequality measures from quintiles estimated using linear interpolation method at nominal level 95\%, each based on 1000 replications and 500 bootstrap repetitions.\strut}
\label{tab2}
\hspace*{-1cm}
\begin{tabular}{ccccccc}
\hline
\multicolumn{1}{c}{\textit{F} } & \multicolumn{1}{c}{\textit{n}} & \multicolumn{4}{c}{\textit{Bootstrap}}& \multicolumn{1}{c}{\textit{Wald-type}}\\
\cline{3-7}
\textit{} & \textit{ } & \textit{Gini} & \textit{Theil} & \textit{Atkinson} & \textit{I} & \textit{I}\\
\hline
 & 50 & 0.788  (0.164) & 0.734  (0.327) & 0.785  (0.129) & 0.947  (0.162) & 0.968  (0.163)\\
Lognormal & 100 & 0.813  (0.119) & 0.761  (0.250) & 0.804  (0.097) & 0.960  (0.112) & 0.965  (0.112)\\
 & 250 & 0.837  (0.075) & 0.720  (0.161) & 0.813  (0.062) & 0.967  (0.069) & 0.962  (0.070)\\
 & 500 & 0.840  (0.054) & 0.650  (0.115) & 0.798  (0.045) & 0.955  (0.048) & 0.956  (0.049) \\
\hline
 & 50 & 0.909  (0.128) & 0.921  (0.151) & 0.911  (0.072) & 0.948  (0.165) & 0.949  (0.164) \\
Singh-Maddala & 100 & 0.925  (0.091) & 0.927  (0.108) & 0.914  (0.052) & 0.933  (0.114) & 0.959  (0.116) \\
 & 250 & 0.940  (0.058) & 0.948  (0.069) & 0.938  (0.034) & 0.933  (0.072) & 0.947  (0.072) \\
 & 500 & 0.946  (0.041) & 0.952  (0.049) & 0.946  (0.024) & 0.941  (0.050) & 0.948  (0.051) \\
\hline
 & 50 & 0.902  (0.128) & 0.886  (0.143) & 0.869  (0.069) & 0.939  (0.169) & 0.946  (0.168)\\
Dagum & 100 & 0.914  (0.093) & 0.902  (0.105) & 0.904  (0.051) & 0.952  (0.117) & 0.951  (0.118)\\
 & 250 & 0.902  (0.059) & 0.878  (0.067) & 0.893  (0.033) & 0.940  (0.073) & 0.948  (0.074) \\
 & 500 & 0.925  (0.042) & 0.891  (0.048) & 0.918  (0.024) & 0.943  (0.052) & 0.954  (0.052) \\
\hline
 & 50 & 0.930  (0.158) & 0.939  (0.285) & 0.931  (0.126) & 0.954  (0.170) & 0.964  (0.170) \\
$\chi^2_2$ & 100 & 0.930  (0.111) & 0.933  (0.204) & 0.930  (0.090) & 0.955  (0.117) & 0.952  (0.118) \\
 & 250 & 0.938  (0.071) & 0.939  (0.131) & 0.939  (0.058) & 0.951  (0.072) & 0.952  (0.073) \\
 & 500 & 0.948  (0.050) & 0.950  (0.093) & 0.946  (0.041) & 0.945  (0.051) & 0.960  (0.051) \\
\hline
 & 50 & 0.633  (0.172) & 0.391  (0.490) & 0.603  (0.177) & 0.968  (0.163) & 0.969  (0.162) \\
Pareto(2) & 100 & 0.637  (0.121) & 0.351  (0.373) & 0.590  (0.131) & 0.970  (0.112) & 0.971  (0.113)\\
 & 250 & 0.571  (0.077) & 0.172  (0.242) & 0.484  (0.084) & 0.949  (0.069) & 0.959  (0.070) \\
 & 500 & 0.500  (0.054) & 0.083  (0.173) & 0.362  (0.060) & 0.973  (0.048) & 0.961  (0.049)\\
\hline
 & 50 & 0.916  (0.158) & 0.934  (0.288) & 0.921  (0.126) & 0.939  (0.169) & 0.965  (0.170) \\
Exponential(1) & 100 & 0.929  (0.111) & 0.938  (0.204) & 0.924  (0.090) & 0.952  (0.116) & 0.966  (0.118) \\
 & 250 & 0.936  (0.071) & 0.949  (0.131) & 0.935  (0.058) & 0.929  (0.072) & 0.962  (0.073) \\
 & 500 & 0.943  (0.050) & 0.945  (0.093) & 0.947  (0.041) & 0.961  (0.050) & 0.963  (0.051) \\
\hline
\end{tabular}
 \end{table}

From Table \ref{tab2} for quintile-grouped data and using the linear interpolation method, intervals for \textit{I} produces coverage probabilities close to the nominal level of 0.95 together with narrow mean width for all settings and with both bootstrap and the Wald-type intervals.  Given that the computation of the interval is much more efficient for the Wald-type interval, there does not appear to be an advantage for using the bootstrap.   However, for the Gini, Theil and Atkinson measures, the coverages are comparatively weaker but improves as the sample size increases for most of the distributions. 

\begin{table}
\small
\renewcommand{\tabcolsep}{1.5mm}
  \centering
  
    \caption{Empirical coverage probabilities and average widths (in brackets) of Boot-strapped interval estimates of inequality measures from quintiles estimated using GLD method at nominal level 95\% for, each based on 1000 replications and 500 bootstrap repetitions.\strut}
\label{tab3}
\hspace*{-1cm}
\begin{tabular}{ccccccc}
\hline
\multicolumn{1}{c}{\textit{F} } & \multicolumn{1}{c}{\textit{n}} & \multicolumn{4}{c}{\textit{Bootstrap}}& \multicolumn{1}{c}{\textit{Wald-type}}\\
\cline{3-7}
\textit{} & \textit{ } & \textit{Gini} & \textit{Theil} & \textit{Atkinson} & \textit{I} & \textit{I}\\
\hline
 & 50  & 0.495  (0.168) & 0.406  (0.387) & 0.598  (0.150) & 0.967  (0.173) & 0.974  (0.172)\\
Lognormal & 100 & 0.446  (0.117) & 0.366  (0.260) & 0.510  (0.101) & 0.975  (0.126) & 0.971  (0.105)\\
& 250 & 0.373  (0.071) & 0.269  (0.141) & 0.453  (0.059) & 0.899  (0.085) & 0.924  (0.065) \\
 & 500 & 0.271  (0.049) & 0.165  (0.090) & 0.359  (0.039) & 0.661  (0.063) & 0.713  (0.046)\\
\hline
 & 50 & 0.862  (0.134) & 0.937  (0.151) & 0.953  (0.090) & 0.979  (0.168) & 0.989  (0.180)\\
Singh-Maddala & 100 & 0.783  (0.094) & 0.920  (0.107) & 0.930  (0.063) & 0.984  (0.119) & 0.973  (0.125)\\
& 250 & 0.735  (0.060) & 0.918  (0.068) & 0.911  (0.040) & 0.974  (0.075) & 0.955  (0.078)\\
& 500 & 0.646  (0.042) & 0.887  (0.048) & 0.803  (0.028) & 0.965  (0.054) & 0.925  (0.056)\\
\hline
& 50 & 0.844  (0.133) & 0.955  (0.140) & 0.988  (0.085) & 0.990  (0.174) & 0.988  (0.192)\\
Dagum & 100 & 0.759  (0.094) & 0.909  (0.099) & 0.991  (0.060) & 0.991  (0.123) & 0.981  (0.132)\\
& 250 & 0.561  (0.060) & 0.799  (0.063) & 0.982  (0.038) & 0.982  (0.079)) & 0.959  (0.083) \\
 & 500 & 0.299  (0.042) & 0.575  (0.045) & 0.981  (0.027) & 0.967  (0.057) & 0.941  (0.059)\\
\hline
& 50 & 0.652  (0.169) & 0.544  (0.359) & 0.749  (0.158) & 0.980  (0.170) & 0.989  (0.172)\\
$\chi^2_2$ & 100 & 0.583  (0.121) & 0.488  (0.269) & 0.663  (0.111) & 0.971  (0.117) & 0.978  (0.118)\\
& 250 & 0.605  (0.073) & 0.512  (0.147) & 0.666  (0.065) & 0.970  (0.073) & 0.979  (0.073)\\
& 500 & 0.568  (0.051) & 0.467  (0.096) & 0.624  (0.044) & 0.974  (0.051) & 0.969  (0.051)\\
\hline
& 50 & 0.558  (0.237) & 0.508  (1.029) & 0.609  (0.289) & 0.973  (0.161) & 0.989  (0.161)\\
Pareto(2) & 100 & 0.579  (0.197) & 0.549  (1.056) & 0.607  (0.251) & 0.971  (0.111) & 0.977  (0.111)\\
& 250 & 0.626  (0.152) & 0.647  (0.982) & 0.663  (0.201) & 0.968  (0.069) & 0.972  (0.069)\\
& 500 & 0.650  (0.123) & 0.697  (0.903) & 0.687  (0.169) & 0.976  (0.048) & 0.977  (0.049)\\
\hline
& 50 & 0.653  (0.172) & 0.559  (0.388) & 0.722  (0.163) & 0.973  (0.169) & 0.980  (0.171)\\
Exponential(1) & 100 & 0.589  (0.119) & 0.513  (0.259) & 0.667  (0.110) & 0.970  (0.117) &  0.983  (0.118)\\
& 250 & 0.578  (0.074) & 0.483  (0.151) & 0.651  (0.066) & 0.982  (0.073) & 0.973  (0.073)\\
& 500 & 0.561  (0.051) & 0.470  (0.095) & 0.615  (0.044) & 0.973  (0.051) & 0.969  (0.051)\\
\hline
\end{tabular}
 \end{table}

Table \ref{tab3} shows that the intervals based on the GLD and quintiles for the Gini, Theil and Atkinson measures have poor coverage.  Coverages are typically very good for the QRI intervals, albeit more conservative than those using the linear interpolation method.  However, coverages become low for the lognormal suggesting that quintiles do not provide enough information to get a good approximation using the GLD.

\begin{table}
\small
\renewcommand{\tabcolsep}{1.5mm}
  \centering
  
    \caption{Empirical coverage probabilities and average widths (in brackets) of Boot-strapped interval estimates of inequality measures from deciles estimated using GLD method at nominal level 95\% for, each based on 1000 replications and 500 bootstrap repetitions.\strut}
\label{tab4}
\hspace*{-1cm}
\begin{tabular}{ccccccc}
\hline
\multicolumn{1}{c}{\textit{F} } & \multicolumn{1}{c}{\textit{n}} & \multicolumn{4}{c}{\textit{Bootstrap}}& \multicolumn{1}{c}{\textit{Wald-type}}\\
\cline{3-7}
\textit{} & \textit{ } & \textit{Gini} & \textit{Theil} & \textit{Atkinson} & \textit{I} & \textit{I}\\
\hline
 & 50 & 0.754  (0.262) & 0.733  (0.963) & 0.762  (0.273) & 0.926  (0.156) & 0.948  (0.156) \\
Lognormal & 100 & 0.789  (0.209) & 0.787  (0.892) & 0.781  (0.227) & 0.943  (0.108) & 0.953  (0.109) \\
 & 250 & 0.760  (0.152) & 0.761  (0.749) & 0.756  (0.173) & 0.938  (0.068) & 0.943  (0.068)\\
 & 500 & 0.740  (0.113) & 0.744  (0.585) & 0.730  (0.130) & 0.927  (0.048) & 0.920  (0.048)\\
\hline
 & 50 & 0.791  (0.148) & 0.760  (0.248) & 0.769  (0.103) & 0.912  (0.160) & 0.958  (0.161)\\
Singh-Maddala & 100 & 0.781  (0.102) & 0.756  (0.167) & 0.747  (0.068) & 0.922  (0.111) & 0.965  (0.113)\\
 & 250 & 0.786  (0.060) & 0.748  (0.083) & 0.715  (0.037) & 0.941  (0.070) & 0.954  (0.071)\\
 & 500 & 0.756  (0.041) & 0.706  (0.052) & 0.660  (0.025) & 0.945  (0.050) & 0.955  (0.050)\\
\hline
 & 50 & 0.735  (0.146) & 0.631  (0.222) & 0.740  (0.101) & 0.898  (0.163) & 0.937  (0.167) \\
Dagum & 100 & 0.744  (0.099) & 0.632  (0.138) & 0.733  (0.067) & 0.941  (0.115) & 0.956  (0.118) \\
 & 250 & 0.709  (0.060) & 0.564  (0.074) & 0.685  (0.039) & 0.957  (0.073) & 0.960  (0.074) \\
 & 500 & 0.710  (0.042) & 0.499  (0.047) & 0.681  (0.027) & 0.957  (0.052) & 0.949  (0.052) \\
\hline
 & 50 & 0.807  (0.202) & 0.783  (0.551) & 0.845  (0.196) & 0.941  (0.165) & 0.958  (0.166)\\
$\chi^2_2$ & 100 & 0.775  (0.141) & 0.736  (0.392) & 0.803  (0.134) & 0.954  (0.115) & 0.952  (0.116)\\
 & 250 & 0.799  (0.084) & 0.763  (0.216) & 0.779  (0.077) & 0.969  (0.071) & 0.959  (0.072)\\
 & 500 & 0.753  (0.057) & 0.714  (0.136) & 0.742  (0.050) & 0.970  (0.050) & 0.957  (0.051)\\
\hline
 & 50 & 0.747  (0.283) & 0.682  (1.374) & 0.775  (0.355) & 0.930  (0.159) & 0.948  (0.160)\\
Pareto(2) & 100 & 0.787  (0.236) & 0.745  (1.414) & 0.800  (0.312) & 0.945  (0.110) & 0.939  (0.111)\\
 & 250 & 0.815  (0.185) & 0.817  (1.370) & 0.839  (0.258) & 0.935  (0.068) & 0.911  (0.069)\\
 & 500 & 0.812  (0.149) & 0.856  (1.244) & 0.845  (0.214) & 0.905  (0.048) & 0.928  (0.048)\\
\hline
 & 50 & 0.802  (0.200) & 0.762  (0.537) & 0.822  (0.192) & 0.920  (0.165) & 0.953  (0.167)\\
Exponential(1) & 100 & 0.830  (0.142) & 0.780  (0.395) & 0.826  (0.135) & 0.943  (0.115) & 0.959  (0.116)\\
 & 250 & 0.785  (0.087) & 0.743  (0.232) & 0.781  (0.080) & 0.968  (0.071) & 0.957  (0.072)\\
 & 500 & 0.756  (0.057) & 0.720  (0.139) & 0.748  (0.051) & 0.972  (0.050) & 0.953  (0.051)\\
\hline
\end{tabular}
 \end{table}

When the data is summarised in deciles rather than quintiles (i.e. more bins and more information), Table \ref{tab4} shows improved coverage is achieved with the GLD method.  However, coverage is still poor for the Gini, Theil and Atkinson measures when compared to the good coverages achieved for the QRI.  Again, the similar coverages for the bootstrap and  Wald-type intervals suggest that the Wald-type is a good choice since it is simple and quick to compute.

 \begin{figure}[h]
\centering
\includegraphics[height=8cm,width=10cm]{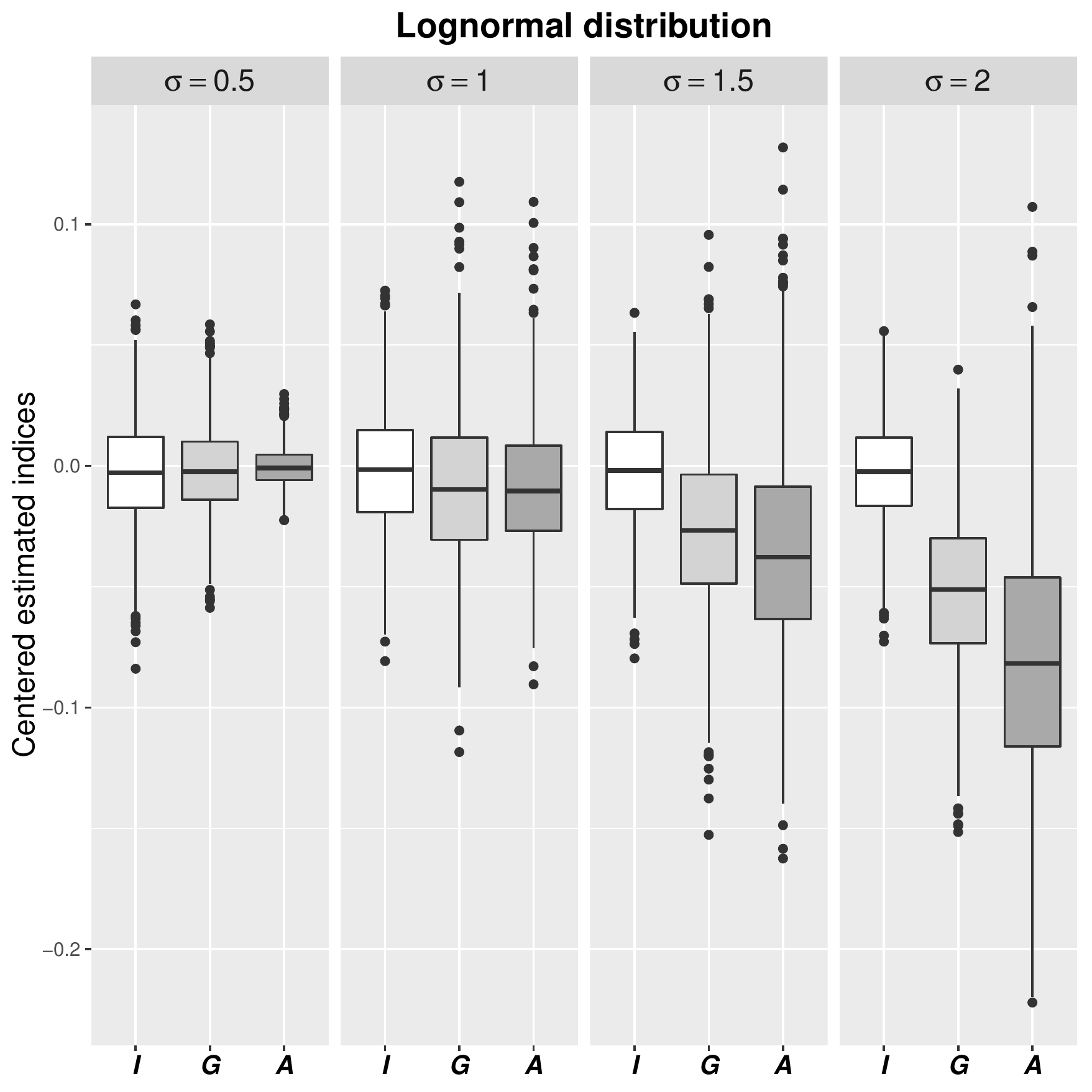}
\caption{Boxplots of 1000 centered (with respect to the true values) simulated estimates of inequality measures from quintiles, estimated using linear interpolation method from the Lognormal distribution with mean 0 and various standard deviation values where \textit{n}=250}
\label{fig1}
\end{figure}

In Figure \ref{fig1} we look at what happens to estimates using the linear interpolation method for each measure (e.g. an estimate based on a bootstrap sample) as skew increases.  In this case, we use the lognormal distribution while increasing the $\sigma$ parameter from 0.5 to 2.  The estimates are centered according to the true value so a value of zero indicates a perfect estimate.  We exclude the Theil index from the analysis since its upper bound is unrestricted. As the distribution becomes more skewed, the Gini and Atkinson estimators have an increase in bias and variability whereas the quantile-based measure (\textit{I})  indicates smaller variability and smaller bias throughout for all of the choices of $\sigma$.  This helps to explain why the coverages are poor for the Gini and Atkinson measures.

\section{Applications} \label{Applications}

\subsection{Example 1: Household income reported with group means} \label{Example 1}

In this example, we present household income data reported with group means by the Survey of Consumer Finances and Expenditures carried out by the Macquarie University and the University of Queensland which can be found in \cite{podder1972distribution} and \cite{kakwani1976efficient}. The data is summarised in Table \ref{tab5}. 

\begin{table}[h!t]
\small
  \centering
    \caption{ Australian household income data for 1967-68 \strut}
\label{tab5}
\hspace*{-1cm}
\begin{tabular}{ccc}
\hline
Income & Number of households & Mean income\\
\hline
Below \$1000 & 310 & 674.39\\
\$1000 - \$2000 & 552 & 1426.10\\
\$2000 - \$3000 & 1007 & 2545.79\\
\$3000 - \$4000 & 1193 & 3469.35\\
\$4000 - \$5000 & 884 & 4470.33\\
\$5000 - \$6000 & 608 & 5446.60\\
\$6000 - \$7000 & 314 & 6460.93\\
\$7000 - \$8000 & 222 & 7459.14\\
\$8000 - \$9000 & 128 & 8456.66\\
\$9000 - \$11000 & 112 & 9788.38\\
\$11000 and over & 110 & 15617.69\\
\hline
\end{tabular}
 \end{table}

\begin{table}[h!t]
\small
  \centering
      \caption{Interval and point estimates of the inequality measures generated using the linear interpolation (LI) and GLD methods for the data presented in Table \ref{tab7}.\strut}
\label{tab6}
\hspace*{-1cm}
\begin{tabular}{cccccc}
\hline
\multirow{2}{*}{\textit{Method}}  & \multicolumn{4}{c}{\textit{Bootstrap}}& \multicolumn{1}{c}{\textit{Wald-type}}\\
\cline{2-6}
& \textit{Gini} & \textit{Theil} & \textit{Atkinson} & \textit{I} & \textit{I} \\
\hline
LI &  0.319 & 0.178 & 0.088 & 0.509 & 0.510 \\
 &  (0.311, 0.327) & (0.168, 0.188) & (0.084, 0.092) & (0.503, 0.517) & (0.502, 0.517)\\
\hline 
GLD & 0.329 & 0.177 & 0.104 & 0.519 &  0.521 \\
 & (0.321, 0.337) & (0.165, 0.190) & (0.098, 0.109) & (0.512, 0.528) & (0.513, 0.529)\\
\hline
\end{tabular}
 \end{table}
 
The confidence intervals produced by 500 bootstrapped samples using the linear interpolation (LI) and GLD methods are given in Table \ref{tab6}. As the final interval is unbounded, we arbitrarily set the upper limit of that bin to \$500,000.  As can be seen, the confidence intervals and the estimates generated by the two methods are similar.

\subsection{Example 2: Comparison of equalized disposable household income data } \label{Example 2}

In this example, we compare two assumed-independent income distributions reported in deciles from \cite{australian2011household} (see Table \ref{tab7}) to assess whether the income inequality measures of the two distributions are significantly different from one another.   It is simple to adapt the previous intervals to the two-sample setting.  For example, for the bootstrap approach we simply estimate the difference at each iteration and then form the interval by taking percentiles from the bootstrapped differences.  For the Wald-type approach we can get the variance of the difference as a sum of the variances for each estimator of the QRI. For estimation purposes, the highest income has been considered as \$5000 for both years. 

\begin{table}
\small
  \centering
  
    \caption{Equalized disposable household income at top of selected percentiles(\$) in Western Australia.\strut}
\label{tab7}
\hspace*{-2cm}
\begin{tabular}{ccc}
\hline
Percentile & 1996-97 & 2009-10\\
\hline
10th & 263 & 347\\
20th & 311 & 454\\
30th & 364 & 565\\
40th & 434 & 663\\
50th & 518 & 770\\
60th & 586 & 882\\
70th & 665 & 1071\\
80th & 778 & 1296\\
90th & 955 & 1652\\
\hline
\end{tabular}
 \end{table}

\begin{table}
\small
  \centering
  
    \caption{Point and interval estimates of inequality measures generated using GLD method for Equalized disposable household income in Western Australia presented in table \ref{tab7} \strut }
\label{tab8}
\hspace*{-2cm}
\begin{tabular}{ccccccc}
\hline
\multicolumn{1}{c}{\textit{Year} } & \multicolumn{1}{c}{} & \multicolumn{4}{c}{\textit{Bootstrap}}& \multicolumn{1}{c}{\textit{Wald-type}}\\
\cline{3-7}
& & \textit{Gini} & \textit{Theil} & \textit{Atkinson} & \textit{I} & \textit{I} \\
\hline
 1996-97& Average Est. & 0.262 & 0.107 & 0.053 & 0.488 & 0.489\\
        & CI & (0.253,0.271) & (0.099,0.115) & (0.049,0.057) & (0.473,0.503) & (0.483,0.496)\\
\hline
2009-10 & Average Est. & 0.326 & 0.174 & 0.083 & 0.538 & 0.538\\
        & CI & (0.318,0.334) & (0.163,0.185) & (0.079,0.088) & (0.528,0.548) & (0.531,0.545)\\
\hline
Difference & Average Est. & 0.064 & 0.067 & 0.030 & 0.050 & 0.049\\
& CI & (0.051,0.077) & (0.054,0.08) & (0.025,0.037) & (0.032,0.07) & (0.040,0.058)\\
\hline
\end{tabular}
 \end{table}

From Table \ref{tab8}, it can be seen that all intervals for the difference in the measures do not include zero.  These intervals then suggest that income inequality has change over the years.  We can conclude that inequality of the equalized disposable household income in Western Australia has been significantly increased from 1996-97 to 2009-10.  

\section{Discussion} \label{Discussion}

To preserve confidentiality, it is common for income data to be summarised in grouped format.  We therefore considered interval estimators for several measures, including the popular Gini index and a newly proposed quantile-based measure, the QRI.  Since grouped data contains bin boundaries and frequencies (and therefore quantile estimates of the data), the QRI is naturally suited to this setting.  We showed that bootstrap intervals and a Wald-type interval, both using estimated densities form the grouped data, had typically excellent coverage (i.e. close to nominal).  The other measures, however, often had intervals with poor coverage.  Further research could include consideration of how to get good approximations to the variances of the Gini, Theil and Atkinson estimators when dealing with grouped data.  This was possible for the QRI since the variance of the estimator can be approximated using the estimated density function.  For the other measures it is not so straightforward.  In summary, when faced with grouped data, if confidence intervals are needed then the QRI is a good option for measuring inequality.

\bibliographystyle{authordate4}
\bibliography{ms}

\begin{thebibliography}{}

\bibitem[\protect\citename{ABS, }2011]{australian2011household}
{\sc ABS}. 2011.
\newblock {\em Household income and income distribution, australian bureau of
  statistics report 6523.0, 2009--10}.

\bibitem[\protect\citename{Allison, }1978]{allison1978measures}
{\sc Allison, Paul~D}. 1978.
\newblock Measures of inequality.
\newblock {\em {Am. Sociol. Rev.}},  865--880.

\bibitem[\protect\citename{Atkinson, }1970]{atkinson1970measurement}
{\sc Atkinson, A.~B.} 1970.
\newblock On the measurement of inequality.
\newblock {\em {J. Econ. Theory}}, {\bf 2}(3), 244--263.

\bibitem[\protect\citename{Biewen \& Jenkins, }2006]{biewen2006variance}
{\sc Biewen, Martin, \& Jenkins, Stephen~P}. 2006.
\newblock Variance estimation for generalized entropy and atkinson inequality
  indices: The complex survey data case.
\newblock {\em {Oxf. Bull. Econ. Stat.}}, {\bf 68}(3), 371--383.

\bibitem[\protect\citename{Damgaard \& Weiner, }2000]{damgaard2000describing}
{\sc Damgaard, Christian, \& Weiner, Jacob}. 2000.
\newblock Describing inequality in plant size or fecundity.
\newblock {\em Ecology}, {\bf 81}(4), 1139--1142.

\bibitem[\protect\citename{Dedduwakumara \& Prendergast,
  }2018]{doi:10.1080/03610918.2018.1499935}
{\sc Dedduwakumara, D.~S., \& Prendergast, L.~A.} 2018.
\newblock Confidence intervals for quantiles from histograms and other grouped
  data.
\newblock {\em {Commun. Stat. Simul. Comput.}}, {\bf 0}(0), 1--14.

\bibitem[\protect\citename{Diciccio \& Efron, }1992]{diciccio1992more}
{\sc Diciccio, Thomas, \& Efron, Bradley}. 1992.
\newblock More accurate confidence intervals in exponential families.
\newblock {\em Biometrika}, {\bf 79}(2), 231--245.

\bibitem[\protect\citename{Dixon {\em et~al.\ }\relax, }1988]{dixon1988erratum}
{\sc Dixon, Philip~M, Weiner, Jacob, Mitchell-Olds, Thomas, \& Woodley,
  Robert}. 1988.
\newblock Erratum to ‘bootstrapping the gini coefficient of inequality’.
\newblock {\em Ecology}, {\bf 69}(4), 1307.

\bibitem[\protect\citename{Efron, }1987]{efron1987better}
{\sc Efron, Bradley}. 1987.
\newblock Better bootstrap confidence intervals.
\newblock {\em Journal of the american statistical association}, {\bf 82}(397),
  171--185.

\bibitem[\protect\citename{Freimer {\em et~al.\ }\relax,
  }1988]{freimer1988study}
{\sc Freimer, M., Kollia, G., Mudholkar, G.~S., \& Lin, C.~T.} 1988.
\newblock A study of the generalized tukey lambda family.
\newblock {\em {Commun. Stat. Theory Methods}}, {\bf 17}(10), 3547--3567.

\bibitem[\protect\citename{Gini, }1914]{gini1914sulla}
{\sc Gini, C.} 1914.
\newblock {\em Sulla misura della concentrazione e della variabilit{\`a} dei
  caratteri}.

\bibitem[\protect\citename{Kakwani \& Podder, }1976]{kakwani1976efficient}
{\sc Kakwani, Nanak~C, \& Podder, Nripesh}. 1976.
\newblock Efficient estimation of the lorenz curve and associated inequality
  measures from grouped observations.
\newblock {\em Econometrica: Journal of the econometric society},  137--148.

\bibitem[\protect\citename{Karian \& Dudewicz, }1999]{karian1999fitting}
{\sc Karian, Z.~A., \& Dudewicz, E.~J.} 1999.
\newblock Fitting the generalized lambda distribution to data: a method based
  on percentiles.
\newblock {\em {Commun. Stat. Simul. Comput.}}, {\bf 28}(3), 793--819.

\bibitem[\protect\citename{Kleiber, }2008]{kleiber2008guide}
{\sc Kleiber, C.} 2008.
\newblock A guide to the dagum distributions.
\newblock {\em Pages  97--117 of:} {\em Modeling income distributions and
  lorenz curves}.
\newblock Springer.

\bibitem[\protect\citename{Lyon {\em et~al.\ }\relax,
  }2016]{lyon2016advantages}
{\sc Lyon, M., Cheung, L.~C., \& Gastwirth, J.~L.} 2016.
\newblock The advantages of using group means in estimating the lorenz curve
  and gini index from grouped data.
\newblock {\em {Am. Stat}}, {\bf 70}(1), 25--32.

\bibitem[\protect\citename{McDonald, }1984]{mcdonald1984some}
{\sc McDonald, J.~B.} 1984.
\newblock Some generalized functions for the size distribution of income.
\newblock {\em Econometrica},  647--663.

\bibitem[\protect\citename{Podder, }1972]{podder1972distribution}
{\sc Podder, N}. 1972.
\newblock Distribution of household income in australia.
\newblock {\em Economic record}, {\bf 48}(2), 181--200.

\bibitem[\protect\citename{Prendergast \& Staudte,
  }2016a]{prendergast2016exploiting}
{\sc Prendergast, L.~A., \& Staudte, R.~G.} 2016a.
\newblock Exploiting the quantile optimality ratio in finding confidence
  intervals for quantiles.
\newblock {\em {STAT}}, {\bf 5}(1), 70--81.

\bibitem[\protect\citename{Prendergast \& Staudte,
  }2016b]{prendergast2016quantile}
{\sc Prendergast, L.~A., \& Staudte, R.~G.} 2016b.
\newblock Quantile versions of the {L}orenz curve.
\newblock {\em {Electron. J. Stat.}}, {\bf 10}(2), 1896--1926.

\bibitem[\protect\citename{Prendergast \& Staudte,
  }2018]{prendergast2016simple}
{\sc Prendergast, L.~A., \& Staudte, R.~G.} 2018.
\newblock A simple and effective inequality measure.
\newblock {\em {Am. Stat.}}, {\bf 72}(4), 328--343.

\bibitem[\protect\citename{Prendergast \& Staudte, }2019]{pr&st2019}
{\sc Prendergast, L.~A., \& Staudte, R.~G.} 2019.
\newblock Decomposing the quantile ratio index with applications to australian
  income and wealth data.
\newblock {\em {Eur. J. Pure Appl. Math.}}
\newblock Accepted to appear 8-June-2019.

\bibitem[\protect\citename{{R Core Team}, }2017]{R}
{\sc {R Core Team}}. 2017.
\newblock {\em R: A language and environment for statistical computing}.
\newblock R Foundation for Statistical Computing, Vienna, Austria.

\bibitem[\protect\citename{Shorrocks, }1980]{shorrocks1980class}
{\sc Shorrocks, Anthony~F}. 1980.
\newblock The class of additively decomposable inequality measures.
\newblock {\em Econometrica},  613--625.

\bibitem[\protect\citename{Su, }2009]{su2009confidence}
{\sc Su, S.} 2009.
\newblock Confidence intervals for quantiles using generalized lambda
  distributions.
\newblock {\em {Comput. Stat. Data Anal.}}, {\bf 53}(9), 3324--3333.

\bibitem[\protect\citename{Tarsitano, }2005]{tarsitano2005estimation}
{\sc Tarsitano, A.} 2005.
\newblock Estimation of the generalized lambda distribution parameters for
  grouped data.
\newblock {\em {Commun. Stat. Theory Methods}}, {\bf 34}(8), 1689--1709.

\bibitem[\protect\citename{Theil, }1967]{theil1967economics}
{\sc Theil, H.} 1967.
\newblock {\em Economics and information theory}.
\newblock Tech. rept.

\bibitem[\protect\citename{Wang, }2015]{wang2015bda}
{\sc Wang, B.} 2015.
\newblock {\em bda: Density estimation for grouped data}.
\newblock R package version 5.1.6.

\bibitem[\protect\citename{Zeileis, }2014]{ineq}
{\sc Zeileis, Achim}. 2014.
\newblock {\em {ineq: Measuring Inequality, Concentration, and Poverty}}.
\newblock R package version 0.2-13.

\end{thebibliography}
\end{document}